\documentclass[10pt,letterpaper]{article}
\usepackage[top=0.85in,left=2.75in,footskip=0.75in]{geometry}

\usepackage{amsmath,amssymb}
\usepackage{float}

\usepackage{changepage}

\usepackage{graphicx}

\usepackage[utf8x]{inputenc}

\usepackage{cite}

\usepackage{nameref,hyperref}
\hypersetup{colorlinks=true, linkcolor=blue,
citecolor=blue, filecolor=blue, urlcolor=blue,}

\usepackage[right]{lineno}

\usepackage{microtype}
\DisableLigatures[f]{encoding = *, family = * }

\usepackage[table]{xcolor}

\usepackage{array}

\newcolumntype{+}{!{\vrule width 2pt}}

\newlength\savedwidth

\raggedright
\usepackage[aboveskip=1pt,labelfont=bf,labelsep=period,justification=raggedright,singlelinecheck=off]{caption}

\makeatletter
\renewcommand{\@biblabel}[1]{\quad#1.}
\makeatother

\usepackage{lastpage,fancyhdr,graphicx}
\usepackage{epstopdf}
\fancyheadoffset[L]{2.25in}
\fancyfootoffset[L]{2.25in}
\begin{document}
\vspace*{0.2in}

\begin{flushleft}
{\Large\bf On limitations of uniplex networks for modeling multiplex contagion}
\newline
\\
Nicholas W. Landry\textsuperscript{1,2,3},
jimi adams\textsuperscript{4,5*}
\\
\bigskip
\textbf{1} Vermont Complex Systems Center, University of Vermont, Burlington, VT, USA
\\
\textbf{2} Department of Mathematics and Statistics, University of Vermont, Burlington, VT, USA\\
\textbf{3} Department of Applied Mathematics, University of Colorado Boulder, Boulder, CO, USA
\\
\textbf{4} Department of Health \& Behavioral Sciences, University of Colorado Denver, Denver, CO, USA
\\
\textbf{5} Institute of Behavioral Science, University of Colorado Boulder, Boulder, CO, USA
\bigskip

N.W.L. curated the data, developed the methodology, analyzed the results, and wrote the article. j.a. conceptualized the project, provided supervision, 
 developed the methodology, analyzed the results, and wrote the article.\\[0.1in]

The authors declare no conflicts of interest.\\[0.1in]

* jimi.adams@ucdenver.edu
\end{flushleft}

\newpage

\section*{Abstract}
Many network contagion processes are inherently multiplex in nature, yet are often reduced to processes on uniplex networks in analytic practice. We therefore examine how data modeling choices can affect the predictions of contagion processes. We demonstrate that multiplex contagion processes are not simply the union of contagion processes over their constituent uniplex networks. We use multiplex network data from two different contexts--- (1) a behavioral network to represent their potential for infectious disease transmission using a ``simple'' epidemiological model, and (2) users from  online social network sites to represent their potential for information spread using a threshold-based ``complex'' contagion process. Our results show that contagion on multiplex data is not captured accurately in models developed from the uniplex networks even when they are combined, and that the nature of the differences between the (combined) uniplex and multiplex results depends on the specific spreading process over these networks.

\section*{Introduction}
Studies of contagion on networks regularly rely on data that only represent one type of relationship at a time. These sorts of data can be helpful for understanding contagion dynamics on relationships of that type. However, when the underlying contagion processes are multiplex, they are less well suited for understanding the ultimate extent of spread over a population and the timescale over which these processes occur.

As an example, consider research aiming to model an epidemic of a particular sexually transmitted infection such as human immunodeficiency virus (HIV). HIV is transmissible via sexual contact, shared needles, and other exposures to blood or other bodily fluids. In network terms, this means that any spread of HIV through the population requires an unbroken chain of susceptible cases exposed to those who are currently infectious \cite{moody_epidemic_2017}. That chain of exposure could be contained within relationships of a particular type (e.g., could be completely composed of sexual contacts), or could combine relationships of multiple types (i.e., sexual and needle-sharing contacts).

\begin{figure}
\centering
\includegraphics[width=.8\linewidth]{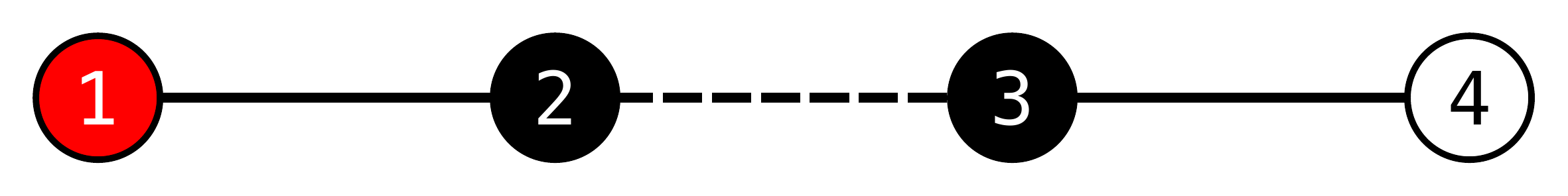}
\caption{{\bf A simple multiplex network.} The line style corresponds with the relationship type. The node color represents its current status: infected (red), susceptible (white), and unknown (black).}
\label{fig:smp}
\end{figure}

In many populations, members have active contacts of both types, and they are not directly overlapping --- for example, not all needle-sharing contacts are with the same partners involving sexual contacts \cite{adams_sex_2013}. In that case, studying only one type of relationship at a time could substantially distort any predictions of epidemic extent within the population. As a simple example, suppose person 1 in Fig.~\ref{fig:smp} is currently infected, and we want to know the likelihood of person 4 becoming infected. If we examine only sexual relationships (denoted by solid lines), both 1 and 4 are sexually active with a single partner (2 and 3, respectively) and those direct relationships would be accurately represented with data on the sexual network along with any models using those data. However, the potential infection-carrying relationship between 2 and 3 providing an indirect route of transmission between 1 and 4 would not be observed, because it stems from a needle-sharing relationship (dashed line). With only data on sexual relationships, we would be underestimating 1's likelihood of \textit{indirectly} infecting 4, and even misunderstanding how these nodes' sexual partnerships contribute to that possible route of transmission. We see that even in this simple example, only considering one tie type can lead to very different predictions of the likelihood of infection and the resulting epidemic extent.

Similarly, there has been a recent proliferation of studies examining propagation of memes, ideas, even voting behavior across social media sites like Facebook or Twitter \cite{aral_distinguishing_2009,bond_61-million-person_2012,lazer_science_2018}. However, while people mix their social media usage and have non-overlapping connectivity patterns across platforms \cite{bode_studying_2018}, these studies overwhelmingly focus on modeling contagion over a single platform at a time. It is reasonable to expect this approach could potentially give a limited picture of the complete contagion potential across the population, as in the sex/needle example for STIs above. Moreover, simply increasing the coverage of the sample used will not overcome these potential sampling biases \cite{hargittai_is_2015}. For example, people may primarily use one platform with their friends and family (ties between $\{1,2\}$ and $\{3,4\}$ in Fig.~\ref{fig:smp}), while using another with their professional contacts (tie $\{2,3\}$ in Fig.~\ref{fig:smp}). In that case, studies that rely on detailed data from a single platform at a time would miss important cross-platform linkages that are vital to understanding the contagion dynamics across the population of interest. Information and misinformation have been shown to spread between different social media platforms, reinforcing the importance of understanding limitations in modeling multiplex contagion \cite{zannettou_web_2017}.

In this paper, we demonstrate the potential limitations of inference about population contagion potential and timescale from networks of a single relationship type (uniplex networks), when contagion processes realistically involve spreading over relationships of multiple types (multiplex networks).

We proceed as follows: first, we motivate the importance of these limitations on contagion modeling; second, we discuss the data sets and contagion processes considered; third, we discuss our approach in quantifying discrepancies in the epidemic trajectories for different choices of data; fourth, we present numerical results validating our premise; and lastly, we discuss limitations and interpretations of this study.

\section*{Motivation}

Our central aim is to show that researchers' ability to model contagion is strongly constrained by choices in the data collection and representation process. Many studies consider a single risk behavior or platform for potential disease/idea spread. This choice can result from simple limitations of data availability, or may draw on the recognition that domain expertise often improves our capacity to collect high quality data on particular topics (e.g., drug sharing networks have considerable barriers to data elicitation \cite{potterat_network_2004}). It is important to understand the limitations introduced by these choices when predicting the extent to which behaviors or diseases will spread. We draw on two simple case studies to explore the discrepancies between the contagion extent resulting from spreading over a single relationship type compared to a model with two different relationship types.

Many studies have presented theoretical approaches for describing contagion dynamics over multiplex networks \cite{de_domenico_physics_2016,gomez_diffusion_2013,radicchi_abrupt_2013,boccaletti_structure_2014}. These studies have been crucial in understanding of dynamics on multiplex networks, but take a fundamentally different approach compared to ours. This study focuses not on predicting the behavior of a multiplex network, but rather exploring the effect of the underlying data representation on the limitations that arise when modeling contagion processes from data representing a single relationship type. This is similar to the approach in Ref.~\cite{kauffman_comparing_2022} where the authors examine how well networks constructed from different data types effectively capture transmission potential. We show that one cannot generate accurate expectations of epidemic extent from uniplex data, even if combining results across layers to reconstruct the multiplex reality, as an estimate of their potential combined effects---and those discrepancies vary by the nature of the contagion process.

\section*{Data}

We use two data sets to represent multiplex networks. For each, we examine (1) the entire multiplex network and (2) the multiplex network constructed with nodes that are members of the largest connected component of the network when considering both relationship types.

The first data set comes from ``Project 90'' in Colorado Springs, which studied a network of commercial sex workers, people who inject drugs, and their sex and needle-sharing partners between 1988-1992 \cite{klovdahl_social_1994}. Here we use the complete network of 7,677 individuals representing all respondents and the `risk behavior' partners they nominated as well as the largest connected component of 4,385 individuals. Although these data were collected over five different waves, we aggregated them into a single network for our purposes. The other data set represents a multiplex network of online social network site interactions among a sample of 1,672 Twitter and Foursquare users \cite{jalili_link_2017} hereafter, `JOAAP' as well as the largest connected component of 1,564 individuals. These data sets are available for download from Refs. \cite{morris_hiv_2011} and \cite{jalili_joapp_2017}, respectively. Summary network properties are presented in Table \ref{tab:network_stats}.

In the Project 90 data, we considered sexual ties as one type of tie and drug or needle-sharing ties as a second tie type. When we refer to the \textit{multiplexed data}, we aggregate these uniplex layers by forming a link between two individuals if they have a sexual relationship or a drug or needle-sharing relationship, whereas the uniplex networks are formed from each tie type separately.

The JOAPP data, which are much higher density, representing Foursquare co-check-ins (frequenting the same locales) as one type of tie and ``follower'' links on Twitter as a second type of tie. We symmetrized the Twitter data for consistent analysis, even though these data are inherently directional.

Analyses begin with the observed multiplex Project 90 and JOAAP data sets \cite{jalili_joapp_2017,klovdahl_social_1994}. We then decompose each of these multiplex networks into their respective uniplex networks with each tie-type representing a unique network. For each data set, this produces three networks. For the Project 90 data set, this produces (1) the multiplex composite risk behavior network, (2a) the sexual contact uniplex network, and (2b) the needle-sharing uniplex network. For the JOAAP data set, this produces (3) the multiplex composite online network, (4a) the Foursquare uniplex network, and (4b) the Twitter uniplex network. All analyses described below are then conducted on each of these networks separately, and the results compared between the multiplex networks (1 and 3) and the union of the uniplex layers (2a and 2b, and 4a and 4b, respectively).

For each data set, we extract the list of node labels by tabulating all the nodes that are connected to either link type. For each type of network (uniplex or multiplex), we start with this node list and then add the links corresponding to the desired tie types. Although in principle one could weight the links according to the frequency of their occurrence, for simplicity, we consider unweighted uniplex and multiplex networks, even if two ties of different types connect the same node pairs, accounting for why the average degree of the two uniplex networks do not sum to the average degree of the multiplexed network.

In our study, we use the Project 90 data as a model of a behavioral network over which epidemics can spread and the JOAPP data as a model of online networks for behavior adoption. Table~\ref{tab:network_stats} presents descriptive statistics for each data set. In Table~\ref{tab:network_stats}, $N$ denotes the number of nodes in the network and $\langle k\rangle$ denotes the average degree. In the susceptible -- infected contagion model, the size of the connected component of which a seed node is a member determines the epidemic size. For this reason, we compute the average size of a connected component given a seed node selected uniformly at random with the expression $\langle c \rangle =\sum_i^{N_c}c_i^2/N$, given connected components $c_1,\dots, c_{N_c}$. Likewise, for complex contagion, the neighborhood of a randomly selected node can be predictive of how behavioral contagion will spread and we compute the average clustering coefficient, $C$, as defined by the authors in Ref.~\cite{watts_collective_1998}. It can be helpful to measure the extent of the overlap between two layers in a multiplex network and there are a wide variety of ways of computing this statistic \cite{kivela_multilayer_2014}. In Table~\ref{tab:network_stats}, we compute a slight modification of the \textit{degree of multiplexity} defined in Ref.~\cite{kapferer_norms_1969} and measure the fraction of links that exist in both layers with respect to the number of links in each of the uniplex networks as well as the multiplexed network. We denote this quantity as $\kappa$.

\begin{table}
\begin{adjustwidth}{-2.25in}{0in}
\caption{{\bf Statistics of the chosen networks} \label{tab:network_stats}}
\begin{tabular}{lccccc}
Data set & $N$ & $\langle k\rangle$ & $\langle c\rangle$ & $C$ & $\kappa$\\
\hline
\textbf{Full data set} &&&&&\\
1. Project 90 & 7677 & 1.70 & 2495.93 & $1.59 \times 10^{-2}$ & $2.13 \times 10^{-2}$\\
2a. Project 90 - Sex & 7677 & 0.85 & 466.44 & $7.78 \times 10^{-4}$ & $4.26 \times 10^{-2}$\\
2b. Project 90 - Drugs & 7677 & 1.17 & 1039.01 & $1.53 \times 10^{-2}$ & $3.08 \times 10^{-2}$\\
3. JOAPP & 1672 & 45.7 & 1672 & 0.519 & $3.32 \times 10^{-2}$\\
4a. JOAPP - Foursquare & 1672 & 35.22 & 1672 & 0.613 & $4.32 \times 10^{-2}$\\
4b. JOAPP - Twitter & 1672 & 16.85 & 1455.58 & 0.123 & $8.99 \times 10^{-2}$\\[0.1in]
\textbf{Largest component of multiplexed data} &&&&\\
1. Project 90 & 4375 & 2.46 & 4375 & $2.62 \times 10^{-2}$ & $1.71 \times 10^{-2}$\\
2a. Project 90 - Sex & 4375 & 1.22 & 816.69 & $1.37 \times 10^{-3}$ & $3.46 \times 10^{-2}$\\
2b. Project 90 - Drugs & 4375 & 1.71 & 819.74 & $2.52 \times 10^{-2}$ & $2.46 \times 10^{-2}$\\
3. JOAPP & 1564 & 36.21 & 1564 & 0.302 & $2.66 \times 10^{-2}$\\
4a. JOAPP - Foursquare & 1564 & 24.67 & 1538.12 & 0.317 & $3.91 \times 10^{-2}$\\
4b. JOAPP - Twitter & 1564 & 18.02 & 1556.02 & 0.131 & $5.35 \times 10^{-2}$\\
\hline
\end{tabular}
\end{adjustwidth}
\end{table}

\section*{Contagion processes}

In this study, we considered two models of contagion: a susceptible -- infected (SI) model representing a epidemiological simple contagion process, and a threshold model representing a complex contagion process, which represent two common model forms of contagion processes.

For the SI model, we define two states: susceptible (S) and infected (I). A susceptible node can be infected by one of its infected neighbors with probability $\beta\Delta t$, where $\beta$ is the infection rate and $\Delta t$ is the interval at which the node states are updated \cite{may_transmission_1987}. In this study, $\Delta t = 1$ week and we update the nodes at the next time synchronously. Once a susceptible node is infected, it will remain infected. Thus, for a connected network, given enough time, every node will eventually become infected.

The threshold model (also known as ``complex contagion'') is a common model for behavior adoption on networks \cite{watts_influentials_2007,valente_network_1995}. For the threshold model, we define, as above, two states: non-adopting (S) and adopting (I). We fix an adoption threshold $\tau$ between 0 and 1. A non-adopting node adopts the opinion if the fraction of its neighbors who have already adopted is larger than $\tau$. Once a node has adopted it will not change its state to become non-adopting. We update the opinions of all the nodes at the next time step synchronously; the threshold model is deterministic once the initial state and threshold value have been specified. Unlike the SI model, a contagion spreading via the threshold process may not reach the entirety of a network, even if it is connected \cite{guilbeault_topological_2021}.

For both models, we select seed nodes uniformly at random to infect and this defines the initial state of our system. We run these contagion processes for a sufficient duration to ensure the epidemic extents have reached equilibrium.

\section*{Approach}

We demonstrate how network data representation choices alter the epidemic extent and rate of contagion spread. The data sets described above are inherently multiplex in nature and serve as case studies in answering our central question.

We start by constructing networks from the data sets as described above. For each data set, we produce the observed multiplex network containing both tie types, and two constituent uniplex networks from each data set.

For each simulation run, we fix a single seed node for each of the four settings of interest. We begin by running these respective contagion processes (the SI model on the Project 90 data set and the threshold model on the JOAPP data set) for the following: (1) the respective multiplex network data, (2) and (3) each of the constituent uniplex layers decomposed from those multiplex networks (as described above), and (4) the union of the infected nodes in each layer of (2) and (3) for each time step in the simulated contagion processes. We present results for the proportion of infected network members at each time step and final epidemic extent. For each setting combination, we generate many realizations of these simulations to form an ensemble of time series.

\section*{Results}

For all contagion processes and each parameter value, we set the time step to a week, i.e., $\Delta t = 1$ and infected a single node at random initially (although we present additional results in the Supporting Information using different numbers of seed nodes). We ran 1000 simulations to form an ensemble for robustness.

For the SI process, we used the following infection rates (which, because $\Delta t=1$, are also infection probabilities): $\beta = 1/75, 1/50, 1/30, 1/20$ (simulations for wider parameter ranges in the Supporting Information provide a robustness check). We simulated up until a time of $t_{max} = 250$ (i.e., 5 years; although we present longer simulations in the Supporting Information as a robustness check).
For the threshold process, we used the following adoption thresholds: $\tau = 1/8, 1/10, 1/12, 1/15$ (simulations for wider parameter ranges in the Supporting Information provide a robustness check). We simulated until $t_{max} = 30$ (although we present longer simulations in the Supporting Information as a robustness check).

\begin{figure}
\begin{adjustwidth}{-2.25in}{0in}
\includegraphics[width=18cm]{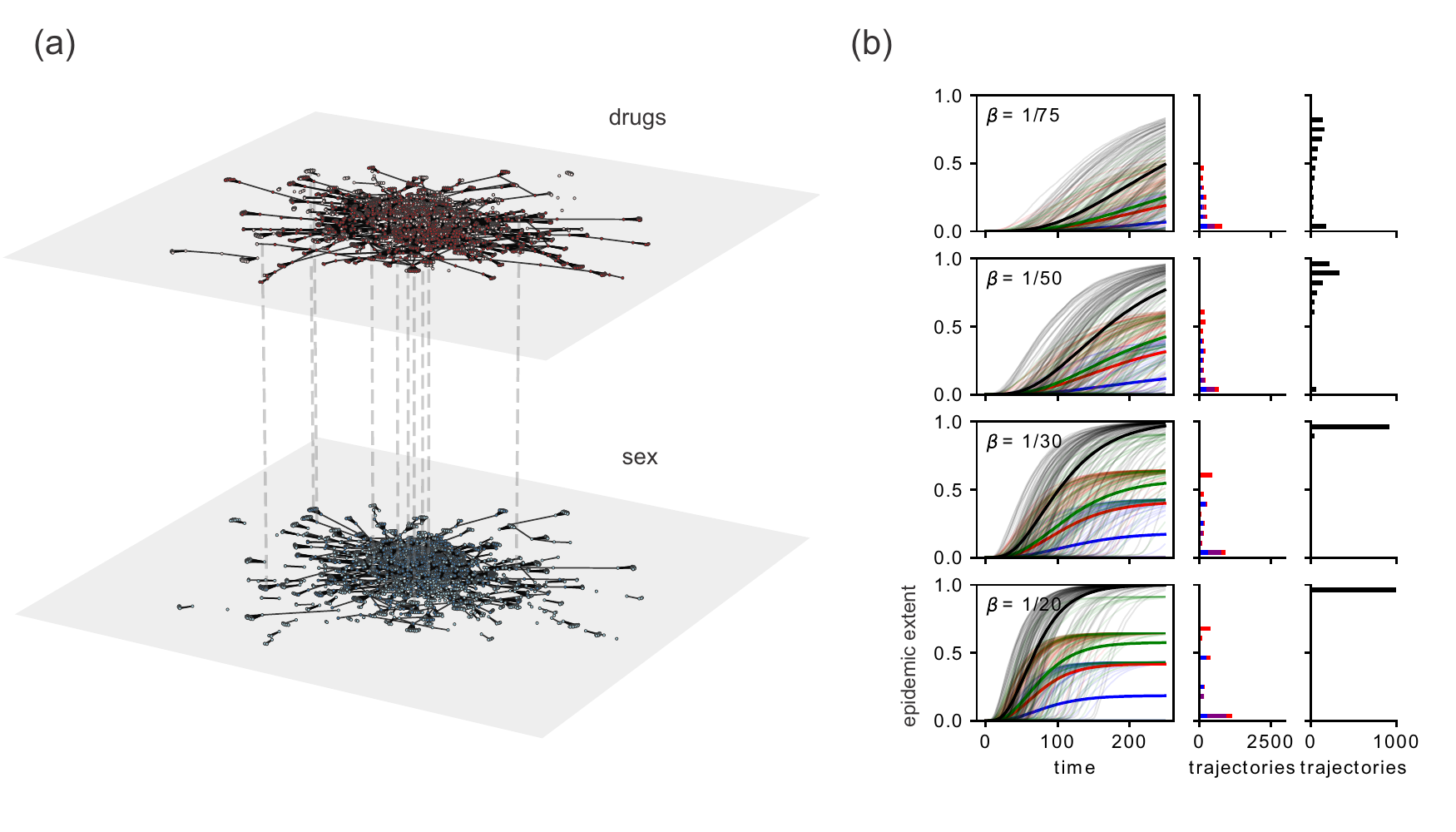}
\caption{{\bf Comparing the epidemic extent for different choices of network data for the Project 90 data set.} (a) A network visualization of the Project 90 data set. The dashed lines are not exhaustive, but illustrate that the nodes in each layer are identical. The lightly colored nodes in each layer are disconnected from the largest connected component in that layer. One can see that each layer in this multiplex network has different contact structure. (b) The SI model simulated on the Project 90 data. The left panel displays a plot of the epidemic extent with respect to time for the uniplex sex network (blue), uniplex drug and needle network (red), the union of uniplex networks (green), and the multiplexed data (black). The thick lines indicate the average epidemic extent and the thin lines are a random sample of 100 individual realizations of this model. The center column illustrates the histogram of epidemic extents for the number of nodes solely accessed from the sex network (blue), the drug network (red), and both (magenta). The right column illustrates the distribution of epidemic extents for the multiplexed data.}
\label{fig:co90}
\end{adjustwidth}
\end{figure}

\begin{figure}
\begin{adjustwidth}{-2.25in}{0in}
\includegraphics[width=18cm]{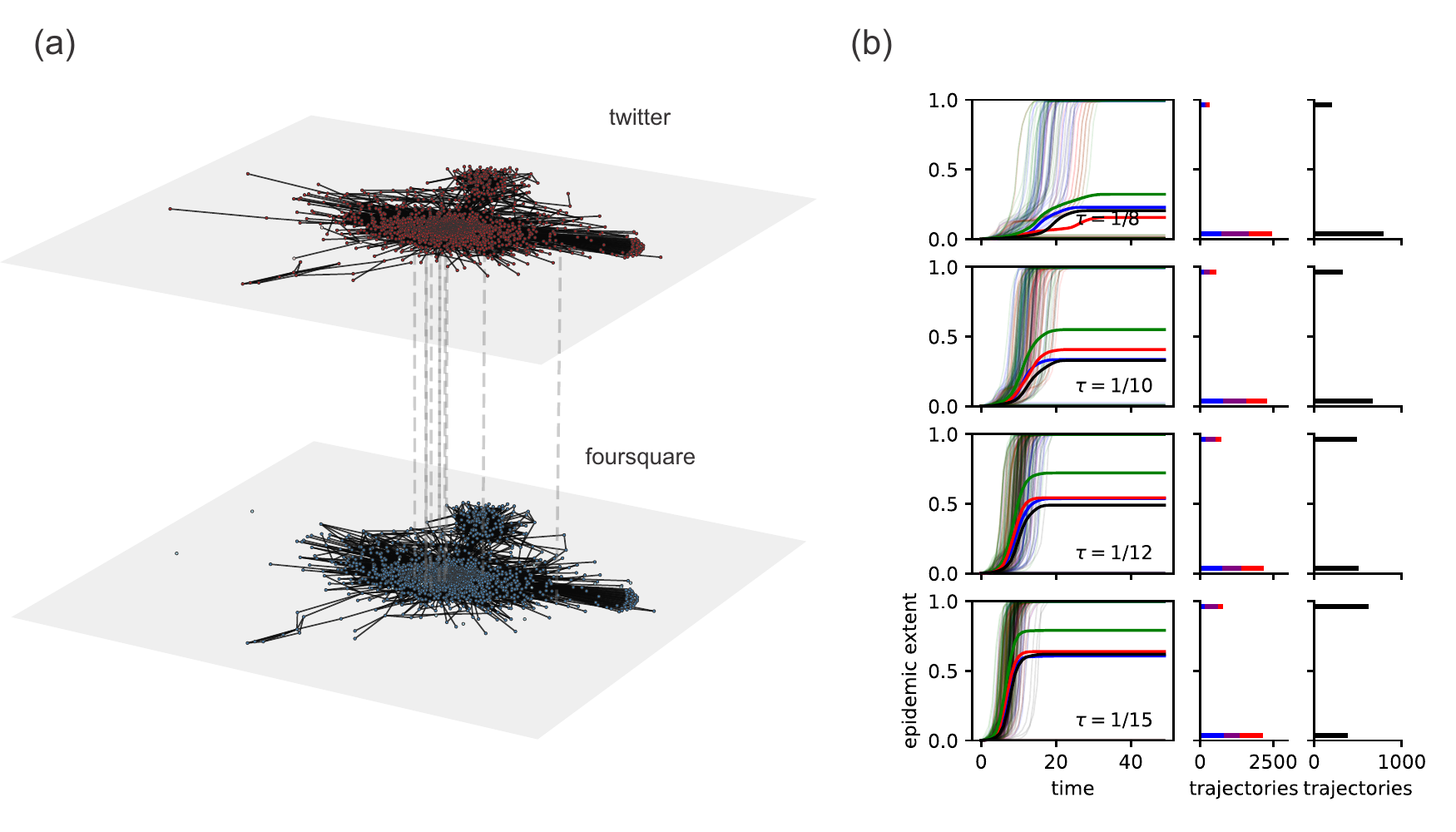}
\caption{{\bf Comparing the epidemic extent for different choices of network data for the JOAPP data set.} (a) Network visualization of the JOAPP data set. The dashed lines are not exhaustive, but illustrate that the nodes in each layer are identical. The light colored nodes in each layer are disconnected from the largest connected component in that layer. (b) The threshold model on the JOAPP data. As in Fig.~\ref{fig:co90}(b) the left column displays a plot of the epidemic extent with respect to time for the uniplex Twitter network (blue), uniplex Foursquare network (red), the union of uniplex networks (green), and the multiplexed data (black). The thick lines indicate the average epidemic extent and the thin lines are a random sample of 100 individual realizations of this model. The center column illustrates the histogram of epidemic extents for the number of nodes solely accessed from the Twitter network (blue), the Foursquare network (red), and both (magenta). The right column illustrates the distribution of epidemic extents for the multiplexed data.}
\label{fig:joapp}
\end{adjustwidth}
\end{figure}

To examine our central question, we draw the reader's attention to comparisons between the black and green curves in Figs.~\ref{fig:co90}(b) and \ref{fig:joapp}(b).  For the SI model, Fig.~\ref{fig:co90}(b) shows that the multiplex data leads to a larger epidemic extent than when we consider the union of the separate uniplex processes.  This result demonstrates that to capture the ``true'' potential of the contagion process on the multiplex network, it is not enough to simply combine the results from their constituent uniplex layers. As shown in Fig.~\ref{fig:smp}, simply considering relationships of one type ignores how contagion can rely on the complementarities across relationships of different types. These uniplex networks, when combined into the multiplex nature of reality, unlock connection patterns that reach more of the population than can be accessed by the combination of the two independently. This indicates, as illustrated in Fig.~\ref{fig:smp}, that there is ``leap-frog'' behavior occurring, where a contagion must pass through connections of alternating types to reach certain nodes.

Looking at the histogram of final extents, we see that for the multiplexed data, by construction, the contagion will spread to the entire population given enough time. The uniplex data sets do not reach the entirety of the population indicating that the uniplex networks have components that are smaller than those of the full multiplex networks.

For the threshold model, the situation is a bit less straightforward. In particular, here the results are more bimodal in that the ultimate contagion extent typically either (a) stalls before taking off at all or (b) reaches the vast majority of the population. This result reflects that both the connectedness and the density of the network determine the epidemic extent (see for example, Ref.~\cite{guilbeault_topological_2021}). This result again demonstrates that understanding contagion processes on uniplex data representation, even when combining them, does not straightforwardly translate to account for contagion on the related multiplex network.
Indeed, in Fig.~\ref{fig:joapp}(b), we see that for each parameter value the epidemic extent for the multiplexed data may be lower than that of both the uniplex network ($\tau = 1/10, 1/12$), larger than that of the Twitter uniplex network but smaller than that of the Foursquare uniplex network ($\tau = 1/15$), and larger than that of the Foursquare uniplex network but smaller than that of the Twitter uniplex network ($\tau = 1/15$). This is likely a complicated interplay between the connectedness and density of these networks. On the one hand, the uniplex networks are more disconnected than the multiplexed network. On the other hand, the multiplexed network is denser than each of the uniplex networks, which discourages the spread of contagion for a given threshold value.  If we set $\tau = 0$, this removes the first of these two factors and the relations between the epidemic extents becomes the same as for the SI model. By definition, the epidemic extent for the union of these is larger and we see that considering both types of relationships separately may be a \textit{worse} estimate than simply considering a single uniplex network (for example $\tau = 1/10, 1/12$ in Fig.~\ref{fig:joapp}(b)).

\section*{Discussion}

In this paper, we have drawn on two cases to show how modeling inherently multiplex contagion processes with uniplex network representations can misrepresent the predicted fraction of the population that a contagion process will reach. We illustrate this limitation separately for models of epidemiological and behavioral contagion to highlight the effect of the data representation on the resulting epidemic behavior. For simplicity, within each model we used the same transmission parameters for each link type. Contagion on empirical networks may have different rates of spread depending on the type of link through which contagion is transmitted (e.g., needle sharing transmits most STIs more efficiently than sexual contacts). In addition, contagion spread can be modeled by different processes spreading on each layer. For example, one might model the spread of awareness as a behavioral process on one network layer and the spread of disease on another network layer \cite{granell_dynamical_2013}.

In the case of the SI model, the multiplex epidemic extent was consistently larger than that of the union of the two constituent uniplex layers. The magnitude of this discrepancy depends on the extent of overlap in the connected components from each uniplex layer. The case where each uniplex network has the same connected components will lead to the same epidemic extent. This is a trivial case where one link type may be a proxy for another link type, which may limit the usefulness of a multiplex representation, although is likely to be uncommon empirically. For the threshold process, gaps remain, but differ in their nature. Whether the union of the constituent uniplex data underestimates or overestimates the true contagion extent depends upon the threshold levels, and these results depend on combinations of network structural characteristics in ways that are important to examine further.

It may be fruitful to estimate how the magnitude of the under or overestimates seen here differ when examining data sets with different structural features. To this end, modeling multiplex data sets using random network models may be helpful in predicting the quantitative differences in the epidemic extent for different data modeling choices.

Our results offer evidence supporting our premise that uniplex data is inadequate for modeling inherently multiplex processes. We offer this study as a cautionary tale for researchers modeling the spread of contagion: the choice of network data is an important assumption baked into contagion models and should be carefully considered. As a more positive recommendation, a primary takeaway of our results is that future data collection efforts should prioritize faithfully capturing the multiplex realities of the underlying processes intended to be examined as in Ref.~\cite{kauffman_comparing_2022}, rather than relying solely on data of a single type.

\section*{Data availability}

The data used in this paper are publicly available from the links provided above. All code used in generating our results is available at \href{https://github.com/nwlandry/multiplex-contagion}{https://github.com/nwlandry/multiplex-contagion} \cite{landry_code_2022}.

\section*{Acknowledgments}
We would like to thank Jeffrey Smith for his helpful feedback on early formulations of this idea, and Ryan Light and David Schaefer for helpful comments on the execution. We would also like to thank Brennan Klein for his \href{https://github.com/jkbren/matplotlib-multilayer-network}{script} for plotting multilayer networks.
N.W.L. acknowledges financial support from the National Science Foundation Grant 2121905, ``HNDS-I: Using Hypergraphs to Study Spreading Processes in Complex Social Networks'', and from the National Institutes of Health 1P20 GM125498-01 Centers of Biomedical Research Excellence Award.

\bibliography{references}

\newpage
\section*{Supporting information for ``On limitations of uniplex networks for modeling multiplex contagion''}

\renewcommand{\thefigure}{S\arabic{figure}}
\setcounter{figure}{0}
\renewcommand{\thetable}{S\arabic{table}}
\setcounter{table}{0}
\renewcommand{\theequation}{S\arabic{equation}}
\setcounter{equation}{0}

\begin{figure}[H]
\begin{adjustwidth}{-2.25in}{0in}
\includegraphics[width=18cm]{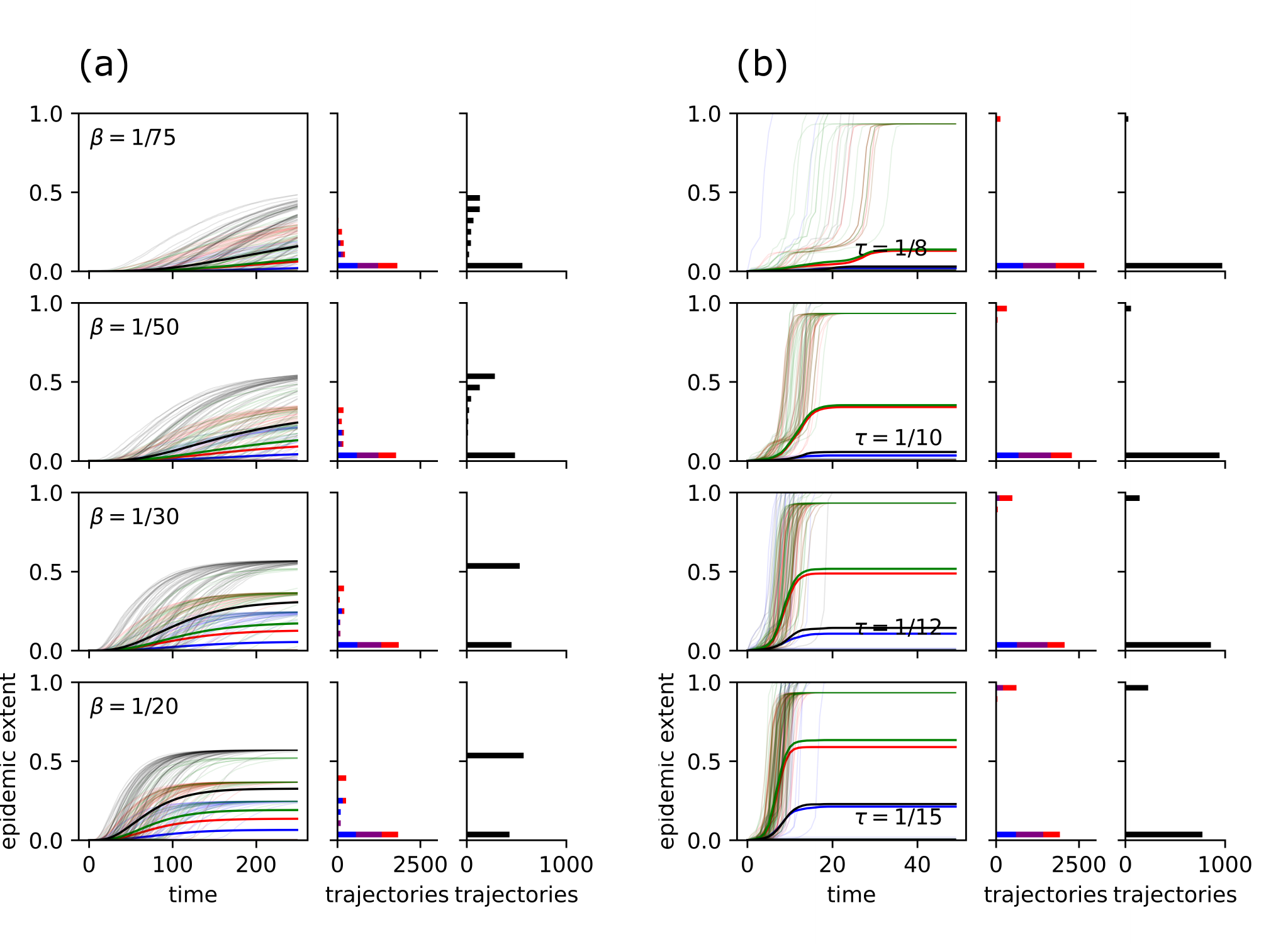}
\caption{}
\label{fig:full_datasets}
\end{adjustwidth}
\end{figure}

\paragraph*{S1 Fig.}
\label{S1_Fig} 
{\bf Full data sets.} We show that our results hold for the full (a) Project 90 and (b) JOAPP data sets, not just the largest component of the multiplexed data. In contrast to Figures 2b and 3b in the main text, the multiplexed data set is no longer fully connected, leading to some epidemic trajectories that reach very few nodes, resulting in the bimodal distribution of epidemic extents. For the Project 90 data set, we see in Fig.~\ref{fig:full_datasets}(a) that the relative epidemic extents are preserved when compared with the epidemic extents of the largest connected component of the multiplexed data. This is not the case with the JOAPP data set in Fig.~\ref{fig:full_datasets}(b), but as discussed in the main text, this is to be expected due to the two competing factors, network density and connectedness, that determine the epidemic extent for the threshold contagion process. For more details on these plots, see Figs. 2b and 3b in the main text.

\newpage

\begin{figure}[H]
\begin{adjustwidth}{-2.25in}{0in}
\includegraphics[width=18cm]{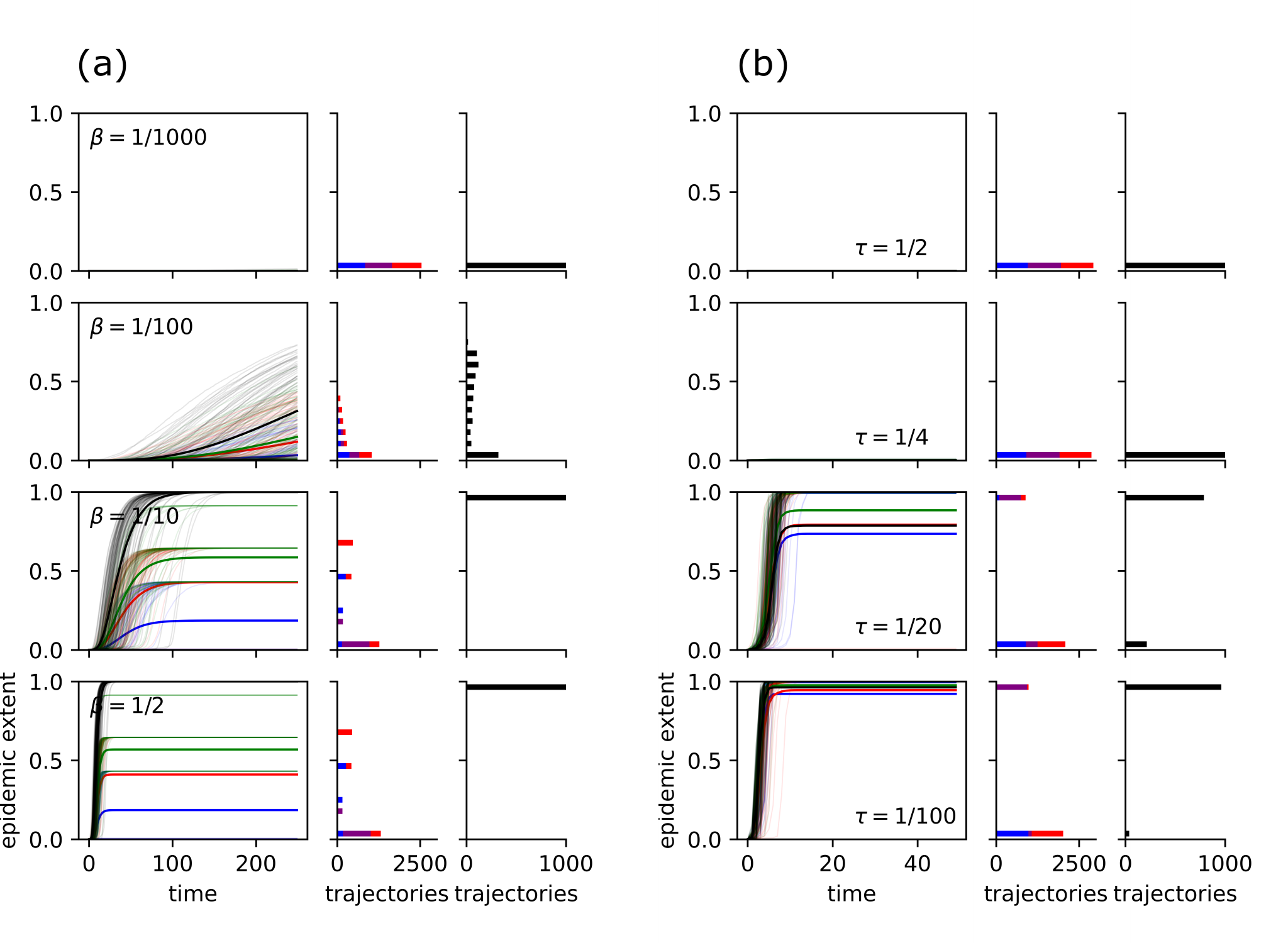}
\caption{}
\label{fig:large_range}
\end{adjustwidth}
\end{figure}

\paragraph*{S2 Fig.}
\label{S2_Fig}
{\bf Larger parameter ranges.} We simulate each contagion process for a wider range of parameters for (a) the Project 90 data set and (b) the JOAPP data set to verify that our results are not dependent on the choice of parameters. For the SI model on the Project 90 data set in Fig.~\ref{fig:large_range}(a), we notice the same differences in epidemic extents, albeit on different time scales. This should be anticipated; if we plot the epidemic extent with respect to $\beta t$, the average of each time series should be the same. For the threshold process on the JOAPP data set in Fig.~\ref{fig:large_range}(b), there are parameter values where the epidemic extents are trivially the same. First, if we choose a threshold greater than the maximum possible fraction of infected neighbors that the network structure and number of seed nodes allow, then contagion will never occur. Second, if the threshold is low enough and the multiplex and uniplex representations are all fully connected, then the entire population will be infected no matter the data representation. For more details on these plots, see Figs. 2b and 3b in the main text.

\newpage

\begin{figure}[H]
\begin{adjustwidth}{-2.25in}{0in}
\includegraphics[width=18cm]{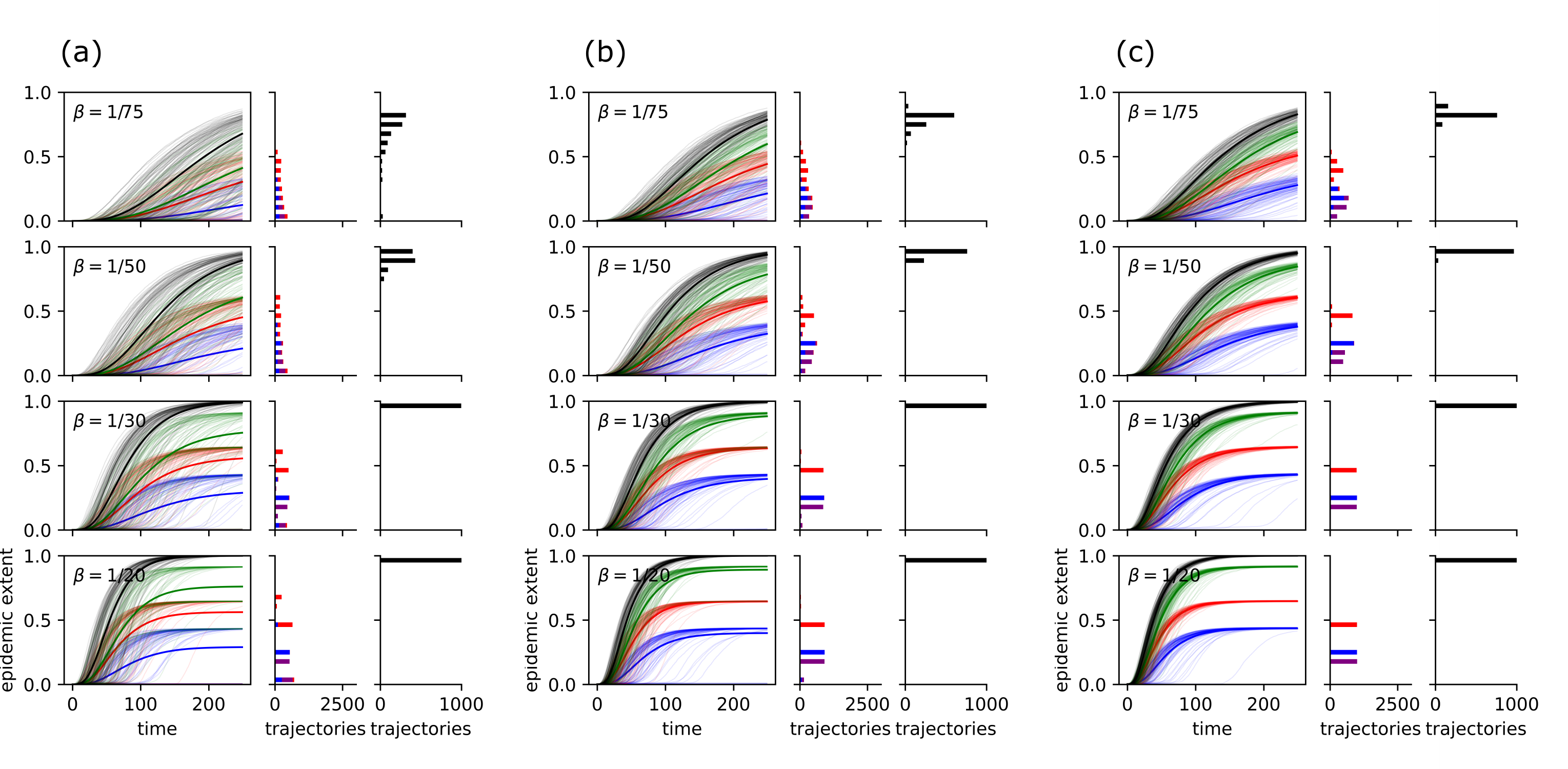}
\caption{}
\label{fig:co90_different_seeds}
\end{adjustwidth}
\end{figure}

\paragraph*{S3 Fig.}
\label{S3_Fig}
{\bf Different number of seeds for the Project 90 data set.} We show that our results hold for (a) 2, (b) 5, and (c) 10 seeds as well as a single seed node as presented in the main text. In Fig.~\ref{fig:co90_different_seeds}, we see that the relative differences in the epidemic extents are preserved. The largest difference is that with a larger number of seed nodes, the likelihood that there is an epidemic trajectory that spreads to very few nodes is much smaller as can be seen in the figure. For more details on these plots, see Figs. 2b and 3b in the main text.

\newpage

\begin{figure}[H]
\begin{adjustwidth}{-2.25in}{0in}
\includegraphics[width=18cm]{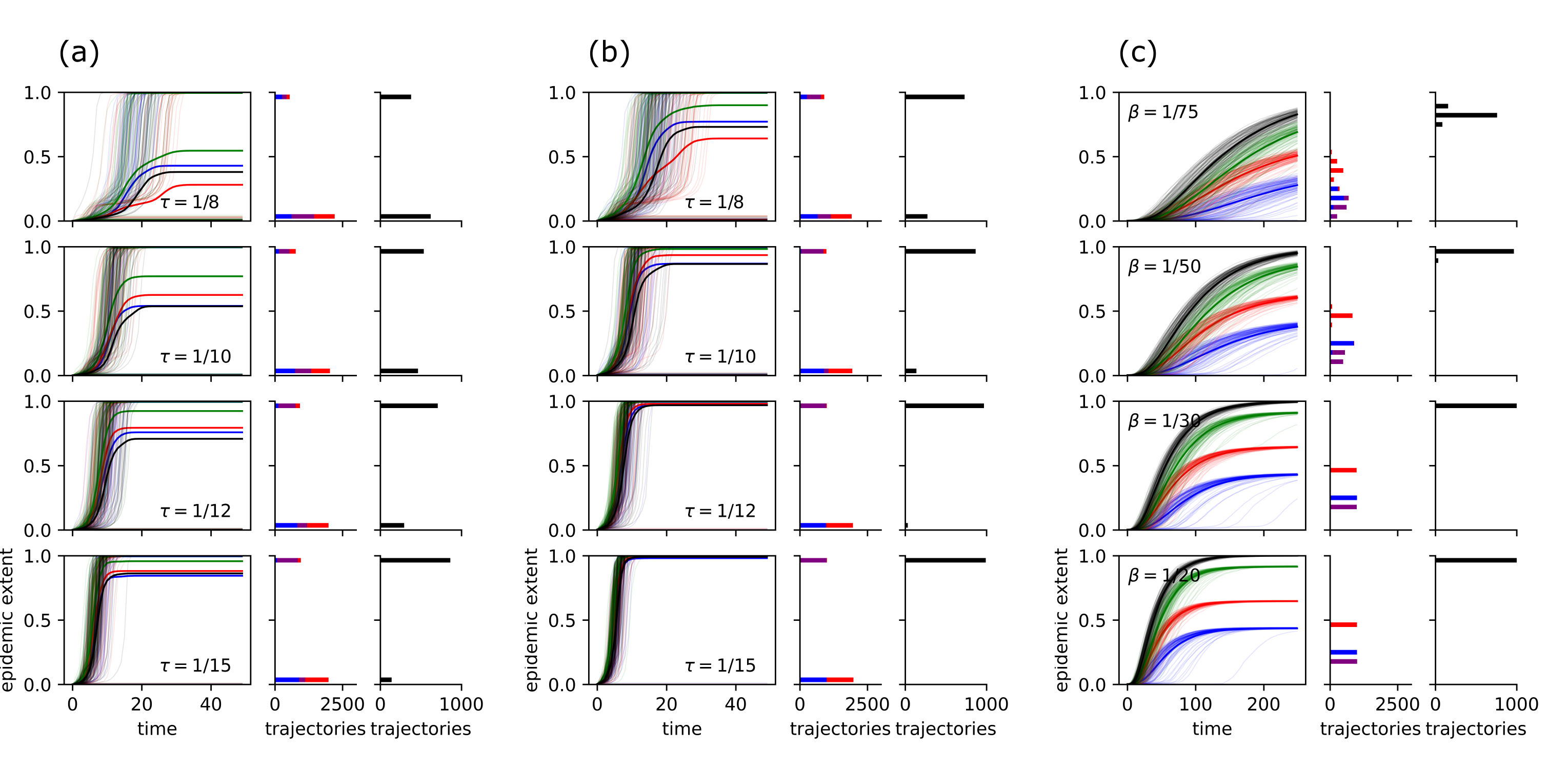}
\caption{}
\label{fig:joapp_different_seeds}
\end{adjustwidth}
\end{figure}

\paragraph*{S4 Fig.}
\label{S4_Fig}
{\bf Different number of seeds for the JOAPP data set.} We show that our results hold for (a) 2, (b) 5, and (c) 10 seeds as well as a single seed node as presented in the main text. In Fig.~\ref{fig:joapp_different_seeds}, we see that, as in Fig.~\ref{fig:co90_different_seeds}, there is a smaller chance of trajectories dying out. In addition, we see that increasing the number of seed nodes effectively raises the maximum threshold for which the contagion will spread to the entire network (for example, $\tau = 1/10$ in Fig.~\ref{fig:joapp_different_seeds}) and lead to trivial results as discussed prior. For more details on these plots, see Figs. 2b and 3b in the main text.

\newpage

\begin{figure}[H]
\begin{adjustwidth}{-2.25in}{0in}
\includegraphics[width=18cm]{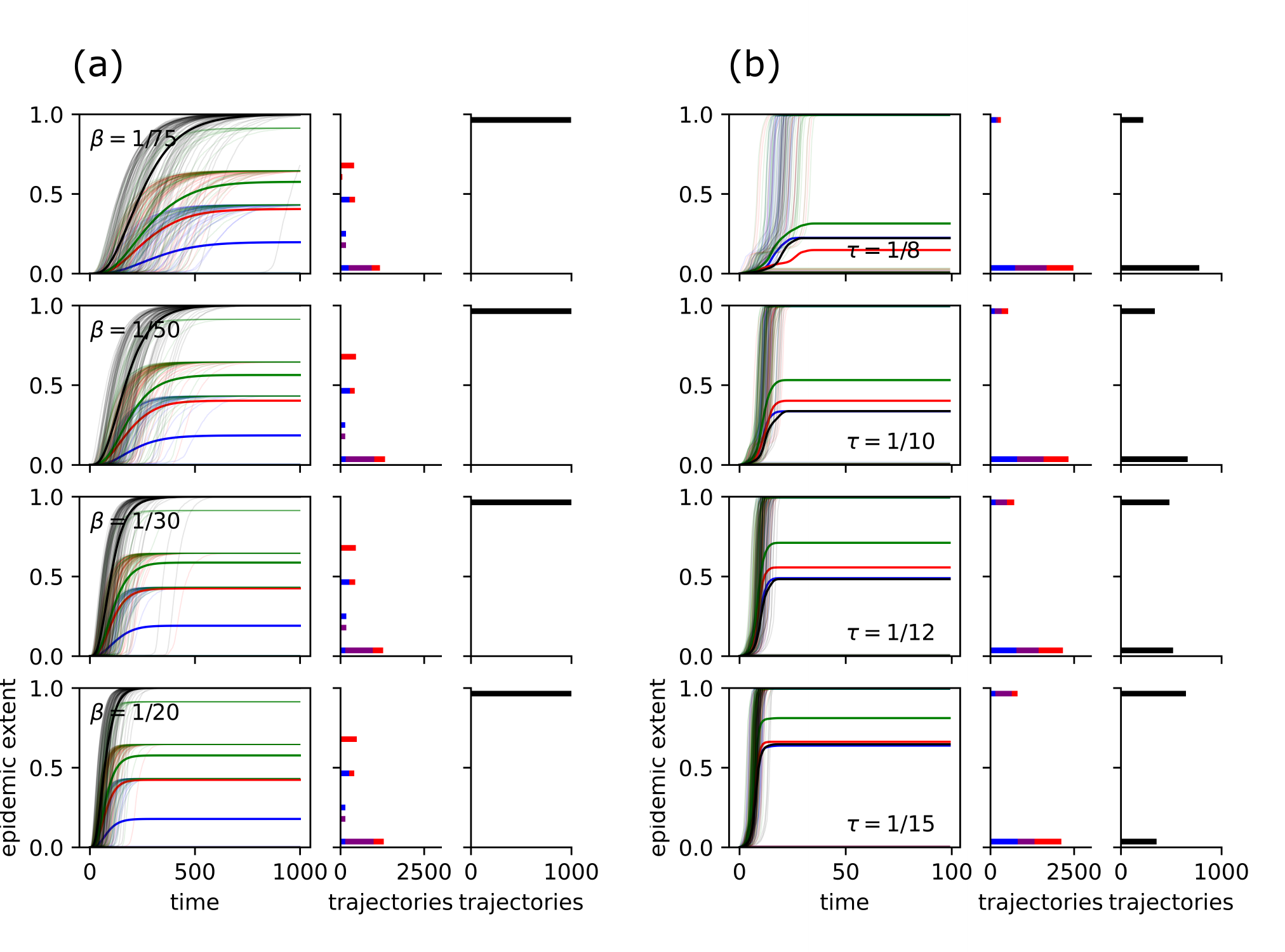}
\caption{}
\label{fig:long_run}
\end{adjustwidth}
\end{figure}

\paragraph*{S5 Fig.}
\label{S5_Fig}
{\bf Full temporal extent.} We run the simulations for a long enough time to remove any temporal censoring for (a) smaller values of $\beta$ for the SI model on the Project 90 data set and for (b) larger values of $\tau$ for the threshold model on the JOAPP data set. In Fig.~\ref{fig:long_run}, we see that the epidemic extents are consistent with our results in the main text. For the SI model, this should be expected as explained prior; rescaling time by the infection probability should yield very similar epidemic responses in expectation. For more details on these plots, see Figs. 2b and 3b in the main text.

\end{document}